\documentclass{amsart}

\usepackage{amsmath,amssymb,amsbsy,amsthm,amsfonts,mathtools,dsfont,hyperref,mathrsfs}

\usepackage{enumitem}
\usepackage{wrapfig}
\usepackage[top=3.5cm,bottom=3.5cm,left=3.5cm,right=3.5cm]{geometry}
\usepackage[normalem]{ulem}

\usepackage{graphicx}
\usepackage{xcolor}
\usepackage{tikz}
\usepackage{tikzscale}
\usetikzlibrary{math}
\usepackage{tikzscale}
\usetikzlibrary{intersections,decorations.markings}
\tikzset{>=latex}
\usepackage{pgfplots}
\pgfplotsset{compat=newest}
\usepgfplotslibrary{fillbetween}
\usepackage[normalem]{ulem}

\allowdisplaybreaks[3]

\newcommand{\tr}{\operatorname{Tr}}

\newcommand\reduline{%
 \bgroup\markoverwith
  {\textcolor{red}{\pgfsetfillopacity{0.2}\rule[-0.5ex]{2pt}{10pt}\pgfsetfillopacity{1}}%
   \textcolor{red}{\llap{\rule[0.4ex]{2pt}{0.4pt}}\llap{\rule[0.7ex]{2pt}{0.4pt}}}%
  }%
  \ULon}

\begin{document}

\newcommand{\ddd}{\,{\rm d}}

\def\os#1{{\color{blue}#1}}
\def\note#1{\marginpar{\small #1}}
\def\tens#1{\pmb{\mathsf{#1}}}
\def\vec#1{\boldsymbol{#1}}
\def\norm#1{\left|\!\left| #1 \right|\!\right|}
\def\fnorm#1{|\!| #1 |\!|}
\def\abs#1{\left| #1 \right|}
\def\ti{\text{I}}
\def\tii{\text{I\!I}}
\def\tiii{\text{I\!I\!I}}

\newcommand{\loc}{{\rm loc}}
\def\diver{\mathop{\mathrm{div}}\nolimits}
\def\grad{\mathop{\mathrm{grad}}\nolimits}
\def\Div{\mathop{\mathrm{Div}}\nolimits}
\def\Grad{\mathop{\mathrm{Grad}}\nolimits}
\def\cof{\mathop{\mathrm{cof}}\nolimits}
\def\det{\mathop{\mathrm{det}}\nolimits}
\def\lin{\mathop{\mathrm{span}}\nolimits}
\def\pr{\noindent \textbf{Proof: }}

\def\pp#1#2{\frac{\partial #1}{\partial #2}}
\def\dd#1#2{\frac{\d #1}{\d #2}}
\def\vec#1{\boldsymbol{#1}}

\def\0{\vec{0}}
\def\A{\mathcal{A}}
\def\B{\mathcal{B}}
\def\b{\vec{b}}
\def\C{\mathcal{C}}
\def\c{\vec{c}}
\def\vv{\vec{v}}
\def\DD{\vec{D}}
\def\Dv{\vec{D}\vv}
\def\BB{\vec{B}}
\def\e{\varepsilon}
\def\er{\epsilon}
\def\f{\vec{f}}
\def\F{\vec{F}}
\def\tF{\tilde{\F}}
\def\g{\vec{g}}
\def\G{\vec{G}}
\def\cG{\mathcal{\G}}
\def\H{\vec{H}}
\def\cH{\mathcal{H}}
\def\I{\vec{I}}
\def\Im{\text{Im}}
\def\j{\vec{j}}
\def\J{\vec{J}}
\def\dd{\vec{d}}
\def\k{\vec{k}}
\def\n{\vec{n}}
\def\q{\vec{q}}
\def\S{\vec{S}}
\def\s{\vec{s}}
\def\T{\vec{T}}
\def\u{\vec{u}}
\def\vp{\vec{\varphi}}
\def\vvt{\vv_{\tau}}
\def\vov{\vv\otimes\vv}
\def\cV{\mathcal{V}}
\def\w{\vec{w}}
\def\W{\vec{W}}
\def\x{\vec{x}}
\def\z{\vec{Z}}
\def\tz{\tilde{\z}}
\def\Z{\vec{Z}}
\def\X{\vec{X}}
\def\Y{\vec{Y}}
\def\balfa{\vec{\alpha}}

\def\Ge{\G_{\e}}
\def\ge{\g_{\e}}
\def\fidv{\phi_{\delta}(|\vv|^2)}
\def\fidve{\phi_{\delta}(|\ve|^2)}
\def\fidvd{\phi_{\delta}(|\vd|^2)}

\def\Ae{\A_\e}
\def\Aee{\Ae^\e}
\def\Aeetilde{\tilde{A}_\e^\e}
\def\Be{\B_\e}
\def\Bee{\Be^\e}
\def\De{\DD\ve}
\def\DDe{\DD^\e}
\def\Dvd{\DD\vd}
\def\oD{\overline{\DD}}
\def\tD{\tilde{\DD}}
\def\Dn{\DD^\e}
\def\Dno{\overline{\Dn}}
\def\Dnt{\tilde{\Dn}}
\def\Dm{\DD^\eta}
\def\Dmo{\overline{\Dm}}
\def\Dmt{\tilde{\Dm}}
\def\Se{\S^\e}
\def\se{\s^\e}
\def\ose{\overline{\se}}
\def\oS{\overline{\S}}
\def\tS{\tilde{\S}}
\def\Sn{\S^\e}
\def\Sno{\overline{\Sn}}
\def\Snt{\tilde{\Sn}}
\def\Sm{\S^\eta}
\def\Smo{\overline{\Sm}}
\def\Smt{\tilde{\Sm}}
\def\ve{\vv^\e}
\def\ove{\overline{\ve}}
\def\vove{\ve\otimes\ve}
\def\vd{\vv^\delta}
\def\sd{\s^\delta}
\def\Sd{\S^\delta}
\def\Dd{\DD\vd}

\def\Wnd#1{W^{1,#1}_{\n, \diver}}
\def\Wndr{W^{1,r}_{\n, \diver}}

\def\o{\Omega}
\def\po{\partial \Omega}
\def\dt{\frac{\d}{\d t}}
\def\pt{\partial_t}
\def\it{\int_0^t \!}
\def\iT{\int_0^T \!}
\def\io{\int_{\o} \!}
\def\iq{\int_{Q} \!}
\def\iqt{\int_{Q^t} \!}
\def\ipo{\int_{\po} \!}
\def\ig{\int_{\Gamma} \!}
\def\igt{\int_{\Gamma^t} \!}

\def\d{\, \textrm{d}}

\def\mn{\mathcal{P}}
\def\du{\mathcal{W}}
\def\tr{\operatorname{tr}}
\def\tow{\rightharpoonup}


\newtheorem{theorem}{Theorem}[section]
\newtheorem{lemma}[theorem]{Lemma}
\newtheorem{proposition}[theorem]{Proposition}
\newtheorem{remark}[theorem]{Remark}
\newtheorem{corollary}[theorem]{Corollary}
\newtheorem{definition}[theorem]{Definition}
\newtheorem{example}[theorem]{Example}

\numberwithin{equation}{section}

\title[On determination of Navier's slip-parameter]{On a methodology to determine Navier's slip-parameter in Navier-Stokes fluid flows at a solid boundary}

\thanks{J.~M\'{a}lek acknowledges the support of the project No. 23-05207S financed by the Czech Science
Foundation (GA \v{C}R). J. M\'{a}lek is a member of the Ne\v{c}as Center for Mathematical Modelling. K~R.~Rajagopal thanks the Office of Naval Research for its support of this work.}

\author[J.~M\'{a}lek]{J.~M\'{a}lek}
\address{Charles University, Faculty of Mathematics and Physics, Mathematical institute, Sokolovsk\'{a} 83, 18675 Prague 8, Czech Republic}
\email{malek@karlin.mff.cuni.cz}

\author[K.~R.~Rajagopal]{K.~R.~Rajagopal}
\address{Department of Mechanical Engineering,  
Texas A\&M University, College Station, TX 77845 USA}
\email{krajagopal@tamu.edu}


\date{\today}

\begin{abstract} 
While the assumption of the ``no-slip" condition at a solid boundary is unquestioningly applied to study the flow characteristics of the Navier-Stokes fluid, there was considerable debate amongst the early pioneers of fluid mechanics, Du Buat, Girard, Navier, Coulomb, Poisson, Prony, Stokes and others, as to the proper condition that pertains at a solid boundary due to a fluid, such as water flowing adjacent to the same. Contemporary usage of the “no-slip” boundary condition notwithstanding, M\'{a}lek and Rajagopal \cite{MKRR2023a} outlined a methodology to test the validity of the assumption. In this study, we continue the investigation further by providing a scheme for determining the slip-parameter that determines the extent of slip, if one presumes that Navier-slip obtains at the boundary. We find that depending on whether the volumetric flow rate is greater or less than the volumetric flow rate corresponding to the no-slip case, different scenarios present themselves regarding what transpires at the boundary.
\end{abstract}

\maketitle

\bigskip

\section{Introduction} \label{Sec1}

While the assumption that the fluid adjacent to an impervious boundary has zero normal component of the velocity with respect to the boundary is physically appropriate, the same cannot be said when it comes to the tangential component of the velocity being zero. The assumption of "no-slip" at a solid impermeable boundary was given the stamp of approval by Stokes \cite{Stokes1845} for sufficiently slow flows in channels and pipes but has become the mainstay with regard to boundary conditions at a solid impermeable boundary. Recently, in \cite{MKRR2023a}, we have evaluated the status of the no-slip boundary condition and provided an experimental procedure  to determine the aptness of the assumption  in the internal  flows of the classical Navier-Stokes fluid. 

We have illustrated our methodology by investigating five unidirectional shear flows of the Navier-Stokes fluid: flows in pipes or between parallel plates (cylindrical and plane Poiseuille and Couette flows), and flows past an inclined plane. Interestingly, the criterion that characterizes the ``no-slip" regime uses (easily) measurable experimental quantities such as the viscosity of the fluid, its density, the pressure gradient (pressure drop), the volumetric flow rate and the geometrical dimensions (such as the radius of the pipe, the distance between the planes, the angle of the inclined plane etc.). 

In this study, we consider the same set of special unidirectional flows as studied in \cite{MKRR2023a}, but we focus on specification of possible flow regimes provided that the criterion for ``no-slip" is not met. Assuming that the conditions (including geometry) of the regimes are unchanged with regard to the five flows studied previously, and that the fluid slips according to Navier's slip (see \cite{navier1823}) that relates the tangential components of the velocity and the normal traction linearly, we are able to determine the exact value of Navier's slip-parameter. The slip-parameter at the solid boundary is however determined only if the volumetric flow rate is greater than the \emph{critical volumetric flow rate}, which is the value of volumetric flow rate corresponding to the ``no-slip" condition. If the volumetric flow rate is less than the critical volumetric flow rate associated with the no-slip boundary condition, one of several scenarios is possible (see comments in the concluding section). Here, we propose a situation similar to that envisaged by Girard, namely, a layer of the fluid near the wall sticks to it and does not deform, that is the layer behaves as though it is rigid layer. Of course, it is possible that the fluid under consideration is not a Navier-Stokes fluid in the first place. 


Let us recall from \cite{MKRR2023a} that Stokes \cite{Stokes1845} was far from convinced that the ``no-slip"  boundary condition for a fluid flowing past a solid boundary at the point of contact was felicitous in general flows. Several of Stokes' forerunners that devoted their research to mechanics, including Du Buat, Coulomb, Girard, Navier, Poisson, Prony and others provided competing assumptions concerning the boundary conditions based on physical considerations, see Goldstein \cite{goldstein1938}. 


The choice of boundary conditions has significant impact on the character of the flows inside the flow domain, see e.g. \cite{Hron2009} for the analytical solutions in a simple geometrical settings and \cite{BMKRR2007, BMKRR2009, BMKRR2020, BMM2023} for advantages that the Navier's slip and other slipping boundary conditions have, in comparison to no-slip boundary condition, on rigorous mathematical properties of long-time and large-data weak solutions of initial- and boundary-value problems concerning (internal) flows of incompressible fluids.


The structure of the paper is as follows. In the following section we develop the governing equations, formulate Navier's slip boundary condition and document that Navier's slip-parameter has to be positive. In Section \ref{Sec3}, we provide a method to determine Navier's slip-parameter for Poiseuille flow in a pipe. We also study the situation when the Navier's slip parameter cannot be determined. In Sections \ref{Sec5}, \ref{Sec6} and \ref{Sec6apul}, we extend the approach to other unidirectional flow problems, namely plane Poisseuille flow, plane Couette flow and cylindrical Couette flow. These flows provide more possible scenarios due to potentially different behavior of the fluid at the separate parts of the solid boundary, i.e., the lower vs upper plate or the inner vs outer cylinder. Finally, in Section \ref{Sec7}, we apply the approach to flows down an inclined plane due to gravity. Section \ref{Sec8} contains concluding remarks. The system of governing equations formulated in the cylindrical coordinates is given in Appendix.

\section{Governing equations and Navier's slip} \label{Sec2}



Our discussion in this paper is concerned with the flows of incompressible Navier-Stokes fluids\footnote{Allowing the possibility that the fluid is non-Newtonian such as for example a power-law fluid, one could give a different flavor to this study. We intend to do this later as we think the understanding of the boundary conditions is somehow more challenging in comparison to the modeling of rheological properties of the fluid in the bulk as boundary conditions depend on the materials on either side of the boundary.} that are characterized by the linear relation between the Cauchy stress tensor and the velocity gradient. 

For the incompressible fluids, the system of governing equations for the unknown velocity field $\vec{v}$ and the pressure (mean normal stress) $p = - \frac13 (\operatorname{tr}\T)$ takes the form  
\begin{align}
    \diver \vec{v} & = 0\,, \label{D0.1} \\
    \varrho \frac{\textrm{d}\vec{v}}{\textrm{d}t} &= \diver \T + \varrho \vec{b}\,, \label{D0.2} \\
    \DD &= \frac{1}{2\mu} (\T +p \I). \qquad\qquad\qquad \Big( p=- \frac13 (\operatorname{tr} \T) \Big) \label{D0.3}
\end{align}
Here, $\varrho$ is the constant (positive) density, $\vec{b}$ is the specific body force, $\mu$ is constant (positive) dynamic (shear) viscosity and $\DD$ stands for the symmetric part of the velocity gradient, i.e., 
\begin{equation}
\DD:=\frac12\left[\nabla \vec{v} + (\nabla \vec{v})^{T}\right].\label{D0.4}
\end{equation}
Note that \eqref{D0.3} is equivalent to the more standard\footnote{The standard way of expressing the Navier-Stokes constitutive relation is at odds with causality as the cause (stress) is expressed in terms of the effect (velocity gradient) (see Rajagopal \cite{Raj.2006} and M\'{a}lek et al. \cite{MPrKRR} for a discussion of the same).} constitutive form for the Navier-Stokes fluid, namely $\T = -p\I + 2\mu \DD$. Note also that the incompressibility constraint \eqref{D0.1} is included automatically in the constitutive equation \eqref{D0.3}, see \cite{Raj.2003,Raj.2006,MPrKRR}.

At the solid impermeable part of the boundary of the flow domains, we will assume, besides the condition of impermeability, that the criterion for ``no-slip" (specified later in the case of particular simple shear flows) is not met and that the fluid slips past the boundary according to the Navier's slip condition. To formulate these conditions we need to fix the notation to be used. Let $\vec{n}$ denote the outward normal vector at a given point of the boundary and $\vec{z}_{\vec{\tau}}$ be the projection of a vector $\vec{z}$ (defined at a point of interest on the boundary) to the tangent plane (constructed at that point of the boundary), i.e. $\vec{z}_{\vec{\tau}}:=\vec{z} - (\vec{z}\cdot\vec{n})\vec{n}$. Then the boundary conditions that we would like to consider take the following form:
\begin{align}
    \vec{v}\cdot\vec{n} & = 0\,, \label{D0.5} \\
    \vec{v}_{\vec{\tau}} & = - \frac{1}{\kappa} (\T\vec{n})_{\vec{\tau}}\,, \label{D0.6} 
\end{align}
where the nonzero slip-parameter $\kappa$ has to be positive as shown next. 

If the flow is internal (i.e. it takes place in a fixed bounded container with no inflows or outflows) and we take a scalar product of $\vec{v}$ and \eqref{D0.2} (where we set $\vec{b} = \vec{0}$ for simplicity), integrate the result over the flow domain $\Omega$,
use the relation $\diver\T\cdot \vec{v} = \diver(\T\vec{v}) - \T \cdot \DD$ together with the Gauss theorem, we obtain, after the integration of the result over the time interval $(0,t)$, that  
\begin{equation}
    \int_{\Omega} \varrho \frac{|\vec{v}(t, \cdot)|^2}{2} + \int_0^t\int_{\Omega}\T\cdot \DD - \int_0^t \int_{\partial \Omega}\T\vec{v} \cdot \vec{n} = \int_{\Omega} \varrho \frac{|\vec{v}(0, \cdot)|^2}{2}.\label{D0.7}
\end{equation}
Noticing that $\T$ is according to \eqref{D0.3} symmetric and \eqref{D0.5} holds, we observe that $\T\vec{v} \cdot \vec{n} = (\T\vec{n})_{\vec{\tau}}\cdot \vec{v}_{\vec{\tau}}$. Hence \eqref{D0.7} leads to 
\begin{equation}
    \int_{\Omega} \varrho \frac{|\vec{v}(t, \cdot)|^2}{2} + \int_0^t\int_{\Omega}\T\cdot \DD + \int_0^t \int_{\partial \Omega}(- \T\vec{n})_{\vec{\tau}} \cdot \vec{v}_{\vec{\tau}} = \int_{\Omega} \varrho \frac{|\vec{v}(0, \cdot)|^2}{2}.\label{D0.8}
\end{equation}
The second and third terms on the left-hand side, considered as a whole, represent the overall dissipation of the system and in accordance with the second-law of the thermodynamics it should be non-negative. As both terms are of different physical nature (the second term corresponds the frictional effects inside the fluid itself, while the third term is due to friction mechanisms on the boundary, i.e. reflecting the properties both of the fluid and the solid), each of them itself is usually assumed to be non-negative, see for example \cite{MKRR2005, MPr2018} and \cite{BMKRR2020} for more details. As a consequence, we observe that the requirement that the second term in \eqref{D0.8} is non-negative together with \eqref{D0.3} imply that $\mu > 0$ and analogously the requirement that the third term is non-negative together with \eqref{D0.6} gives that $\kappa> 0$. Note that letting (formally) $\kappa$ tend to infinity in \eqref{D0.6}, we obtain the ``no-slip" condition 
\begin{equation}
    \vec{v} = \vec{0}. \label{D0.9}
\end{equation}
The fact that $\kappa$ has to be positive will play a key role in our consideration below. 

\section{Poiseuille flow in a pipe} \label{Sec3}

We study the flow of a Navier-Stokes fluid in a cylindrical pipe of infinite length of radius $R$. A constant flow rate $Q$ is driven by a constant negative pressure gradient $c$. We assume that $\vec{b}= \vec{0}$ (no gravity) and $\vec{v} = v(r) \vec{e}_z$. Then, the symmetric part of the velocity gradient $\DD$ simplifies to (see \eqref{A.2})
$$
  \DD = \begin{pmatrix}  0 & 0 & \frac12 \frac{\textrm{d}v}{\textrm{d}r} \\ 
  0&0&0 \\ \frac12 \frac{\textrm{d}v}{\textrm{d}r} & 0& 0\end{pmatrix}\,.
$$
Consequently, the constitutive equation \eqref{D0.3} (see also \eqref{A.3}), takes the reduced form 
\begin{equation}
\tau = \mu \frac{\textrm{d}v}{\textrm{d}r}  \qquad  \textrm{ where }  \tau(r):= T_{rz}(r)\label{au.1}. 
\end{equation}
The governing equations \eqref{D0.2}, see also \eqref{A.1}, then reduce to 
$$
  \frac{\partial p}{\partial r} = \frac{\partial p}{\partial \varphi} = 0 \quad \implies \quad p=p(z)
$$
and 
$$   
   \frac{\textrm{d}p}{\textrm{d}z} = \frac{1}{r} \frac{\textrm{d}(r\tau)}{\textrm{d}r} \qquad \textrm{ where } p=p(z) \textrm{ and } \tau = \tau(r).
$$
Hence 
$$
   \frac{\textrm{d}p}{\textrm{d}z} = c \qquad \textrm{ and } \qquad \frac{\textrm{d}(r\tau)}{\textrm{d}r} = r c.
$$
Consequently,
\begin{equation}
   r \tau = \frac{c r^2}{2} + c_1 \quad \implies \quad \tau = \tau(r) = \frac{cr}{2} + \frac{c_1}{r}\,. \label{CP.0}
\end{equation}
From the constitutive equation \eqref{au.1}, we then obtain 
that 
\begin{equation}
    v(r) = \frac{cr^2}{4\mu} + c_1 \ln r + c_2\,. \qquad (c_1,c_2 \textrm{ are constants})\label{p8}
\end{equation}
The requirement that the velocity is bounded at $r=0$ implies that $c_1 = 0$. 

Let $Q$ be the volumetric flow rate, i.e. $Q = \int_0^R 2\pi r v(r) \, \textrm{d}r$. Using \eqref{p8} with $c_1=0$ we obtain
$$
  Q = \frac{\pi c R^4}{8\mu} + c_2 \pi R^2\,.
$$
Hence 
$$
   c_2 = \frac{1}{\pi R^2} \left[ Q - \frac{\pi c R^4}{8\mu}\right] \quad \textrm{ and } \quad v(r) =  \frac{cr^2}{4\mu} + \frac{Q}{\pi R^2} - \frac{cR^2}{8\mu}\,.
$$
Subtracting and adding the term $cR^2/(4\mu)$ we get
\begin{equation}
  v(r) =  - \frac{c}{4\mu}(R^2 - r^2) + \left[\frac{Q}{\pi R^2} + \frac{c}{8\mu} R^2\right] \,.\label{CP.1}
\end{equation}
If $Q = - \frac{c \pi R^4}{8\mu}$, then $v(R) = 0$ and there is ``no-slip". Note that the quantities $\mu$, $R$, $c$ (the pressure drop) and $Q$ can be (easily) measured.

If $Q\neq - \frac{c\pi R^4}{8\mu}$, then it follows from \eqref{CP.1} (if the assumptions that the flow is unidirectional, steady, and that the pressure is continuous (constant) across the tube etc. are appropriate) that $v(R) \neq 0$ and there is slip. Assuming that the fluid slips past the wall according to the Navier's slip condition \eqref{D0.6} we will be able to show that such a situation is possible only if
\begin{equation}
    Q > - \frac{c\pi R^4}{8\mu}.\label{CP.2}
\end{equation}
Indeed, in the studied geometrical setting, the impermeability condition \eqref{D0.5} is met 
and \eqref{D0.6} reduces to 
\begin{equation}
    v(R) = - \frac{1}{\kappa} \tau(R). \label{CP.3}
\end{equation}
Referring to \eqref{CP.0} (with $c_1 = 0$) and \eqref{CP.1} with $r=R$, the condition \eqref{CP.3} leads to 
\begin{equation}
    \kappa = \frac12 \frac{-cR}{\frac{Q}{\pi R^2} + \frac{c}{8\mu} R^2}. \label{CP.4}
\end{equation}
As $\kappa$ and $-cR$ have to be positive, we see that we can specify $\kappa$ by \eqref{CP.4} only if the denominator in the fraction of \eqref{CP.4} is positive, i.e. if \eqref{CP.2} holds. Said differently, if 
\begin{equation}
    Q < - \frac{c\pi R^4}{8\mu},\label{CP.5}
\end{equation}
then Navier's slip condition \eqref{D0.6} and the solution of the form \eqref{CP.1} are not compatible. Below, following the ideas of Girard \cite{girard1815,Darrigol2002}, we provide one possible scenario of how the solution could look like in the case when the pressure gradient $c$ and the volumetric flow rate $Q$ satisfies  \eqref{CP.5}. 

Let us assume that \eqref{CP.5} holds. This also means that $Q$ and $c$ (as well as $\mu $ and $R$) are fixed. Let us consider $\overline{R}\in (0, R)$ that will be specified later. We look for the flow that is at the rest near the wall in the region characterized by the condition $r\in (\overline{R}, R)$ and that solves the unidirectional shear flow problem with ``no-slip" condition for $r$ satisfying $0\le r < \overline{R}$. It means that for $r$ satisfying $0\le r < \overline{R}$, the solution is of the form (compare with \eqref{CP.1}) 
\begin{equation}
  v(r) =  - \frac{c}{4\mu}(\overline{R}^2 - r^2) \,.\label{CP.7}
\end{equation}
We will now determine $\overline{R}$ by requiring that 
$$
  Q = 2\pi\int_0^{\overline{R}} r v(r) \, dr.
$$
This leads to 
\begin{equation}
  Q = -\frac{c\pi}{8\mu} \overline{R}^4 \quad \implies \quad \overline{R} = \sqrt[4]{\frac{8\mu Q}{-c\pi}}.\label{CP.8}
\end{equation}
It is easy to check that the condition $\overline{R}<R$ coincides with \eqref{CP.5}. 

\begin{figure}[ht]
\input{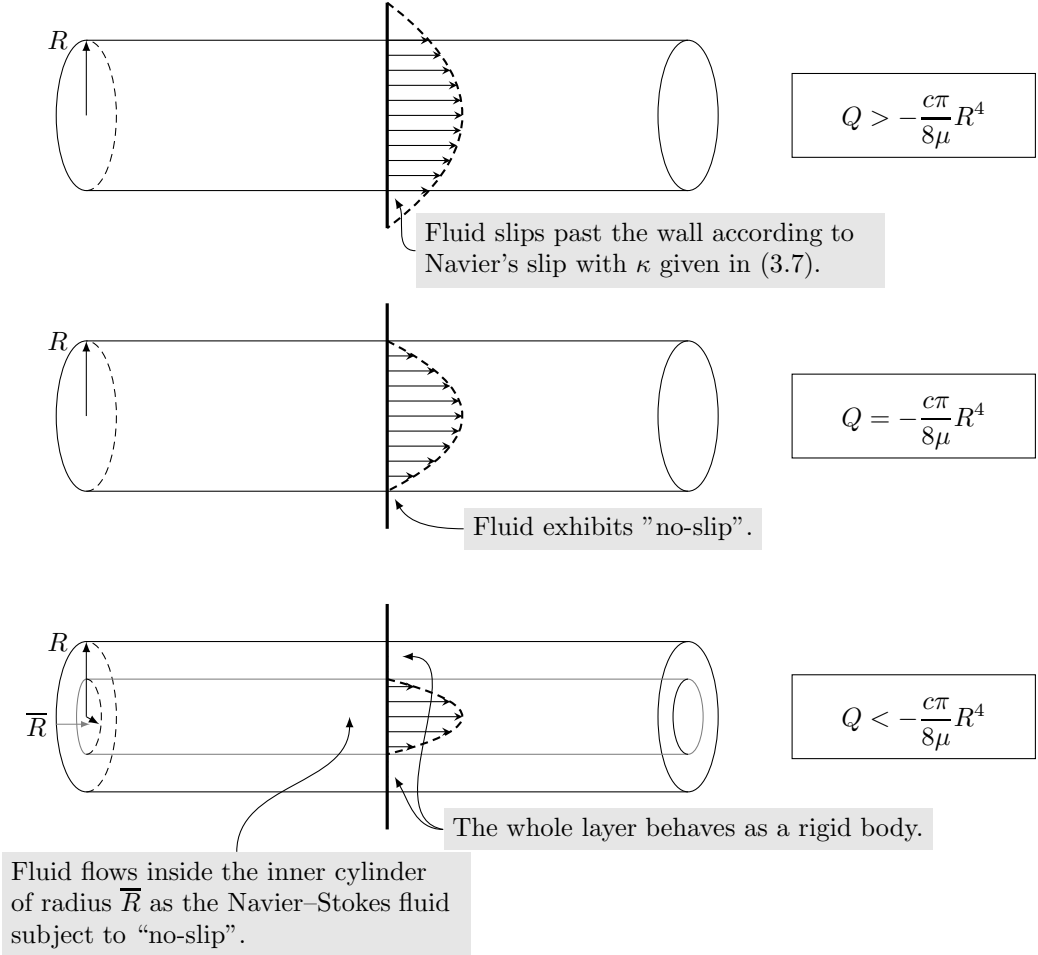}
\begin{tikzpicture}
\pipe{0}{5}{1}
\pipe{40}{0}{2}
\pipe{80}{-5}{3}
\end{tikzpicture}   
\caption{Different flow regimes for Poiseuille flows in a pipe. These regimes depend on the relationship among (easily) measurable quantities $\mu$, $R$, $Q$ and $-c$.}
\label{fig1}
\end{figure}

To conclude, assuming that an incompressible Navier-Stokes fluid
with the shear viscosity $\mu$ flows in a pipe of radius $R$ as a steady unidirectional simple shear flow with a given volumetric flow rate $Q$ and the pressure gradient $-c$ we investigate the case when the ``no-slip" condition 
$$
  Q = -\frac{c\pi}{8\mu} {R}^4
$$
is \emph{not} fulfilled. We have found that 
\begin{itemize}
    \item If $Q> -\frac{c\pi}{8\mu} {R}^4$ and if we assume that the fluid slips past the boundary according to Navier's slip boundary conditions, we have been able to specify the exact value of Navier's slip-parameter $\kappa$, see the formula 
    \eqref{CP.4}. 
    \item If $Q< -\frac{c\pi}{8\mu} {R}^4$, the fluid cannot meet the Navier's slip boundary condition (as otherwise Navier's slip-parameter $\kappa$ would be negative, which contradicts the second law of thermodynamics and the fluid would flow near the wall in the opposite direction as in the inner part of the pipe). Such a solution does not seem to be in keeping with physics.
    \item If $Q< -\frac{c\pi}{8\mu} {R}^4$, the following scenario is possible: the fluid near the outer cylinder is ``stuck" to the wall and the whole layer of the thickness $R-\overline{R}$ together with the wall behaves as a rigid body.  In the inner region when $r$ satisfies $0\le r <  \overline{R}$ the fluid flows according to the formula \eqref{CP.7}. The precise value of $\overline{R}$ is determined from the given data $\mu$, $R$, $Q$ and $-c$ (and the natural assumption that the fluid exhibits ``no-slip" when $r=\overline{R}$), see the formula \eqref{CP.8}.
\end{itemize}
The behavior described above seems plausible, see also Fig. \ref{fig1} for such a viewpoint.

\section{Plane Poiseuille flow} \label{Sec5}

In this case, the flow of a Navier-Stokes fluid is supposed to take place between two parallel plates located at $y=0$ and $y=h$. The effect due to gravity is neglected, i.e. $\vec{b}= \vec{0}$. The flow with a flow rate $Q$ is driven by the constant pressure gradient $c$ that is negative. Also, we assume that $\vec{v} = u(y)\vec{i}$ which gives a very simple form for $\DD$ and this together with the constitutive equation \eqref{D0.3} implies that 
\begin{equation}
\T (x,y,z) = \begin{pmatrix} -p(x,y,z) & \tau(y) & 0 \\
\tau(y) & -p(x,y,z) & 0 \\ 0 & 0 & -p(x,y,z) \end{pmatrix}, \qquad \textrm{ where } \tau := T_{xy}.\label{dod5}
\end{equation}
Then the second and third equation in \eqref{D0.2} lead to $p = p(x)$, while the first equation of \eqref{D0.2} leads to 
$$
   \frac{\textrm{d}p}{\textrm{d}x} = \frac{\textrm{d}\tau}{\textrm{d}y}.
$$
Consequently, as $p=p(x)$ and $\tau = \tau(y)$,
\begin{equation}
    p(x) = cx + b \quad \textrm{ and } \quad \tau(y) = cy + d\,, \label{au.6}   
\end{equation}
where $b$ and $d$ are arbitrary constants, while $c$ is negative constant representing the pressure gradient. The constitutive equation $\frac{\textrm{d}u}{\textrm{d}y} = \frac{\tau}{\mu}$ then leads to 
\begin{equation}
   u(y) = \frac{c}{2\mu} y^2 + \frac{d}{\mu} y + e\,.\label{p2}
\end{equation}
In what follows we assume that the both plates are made of the same material and the fluid at both upper and lower plate interacts in a like manner. It is possible that the interaction at the top and bottom plates could be different, and example of the same is the consequence of the roughness of the two plates being different or one of the plates being hydrophobic and the other hydrophilic and the fluid in question being water. We discuss such a possibility in the next section when analyzing the plane Couette flow.

Under the given situation when the plate-fluid responses are identical at the upper and lower plates, it is reasonable to assume that the velocity profile will be symmetric with respect to the line $y=h/2$. Then it is natural to assume\footnote{We could alternatively impose the condition $u'(h/2) =0$.} that  
\begin{equation}
   u(h) = u(0). \label{p3} 
\end{equation}
Applying this condition to the formula given in \eqref{p2}, we observe that 
\begin{equation}
      \frac{c}{2\mu} h^2 + \frac{d}{\mu} h + e = e \quad \implies \quad d = - \frac{ch}{2}. \label{au.7}   
\end{equation}
Substituting this in \eqref{p2} and requiring further that 
\begin{equation}
   \int_0^h u(y) = Q, \label{p3a} 
\end{equation}
we obtain
$$
  Q = -\frac{c}{12\mu} h^3 + eh \quad \implies \quad e = \frac{Q}{h} + \frac{ch^2}{12\mu}\,.
$$
Hence 
\begin{equation}
    u(y) = \frac{c}{2\mu} y^2 - \frac{ch}{2\mu} y + \frac{ch^2}{12\mu} + \frac{Q}{h}\,.
\label{au.5}
\end{equation}
This implies that 
$$
  u(0) = u(h) =  \frac{ch^2}{12\mu} + \frac{Q}{h}\,.
$$
If $Q = -\frac{ch^3}{12\mu}$, then $u(0)=0$, $u(h)=0$ and the solution of the form 
\begin{equation}
   u(y) = - \frac{c}{2\mu} (h-y) y \label{au.8}
\end{equation}
exhibits ``no-slip" on the lower and the upper plates. If $Q \neq -\frac{ch^3}{12\mu}$, then it follows from \eqref{au.5} that $u(0) \neq 0$ and $u(h) \neq 0$ as well. If $Q > -\frac{ch^3}{12\mu}$, then we will be able to determine (positive) Navier's slip-parameter $\kappa$ appearing in \eqref{D0.6}. Indeed, assuming that the fluid slips via Navier's slip condition \eqref{D0.6}, which in the considered geometrical setting takes the form 
$$
   u(0) = \frac{\tau(0)}{\kappa} \qquad \textrm{ and } \qquad u(h) = - \frac{\tau(h)}{\kappa},
$$
then, using these conditions together with \eqref{au.6}, \eqref{au.7} and \eqref{au.5}, we obtain 
$$
  \kappa = \frac{-6c\mu h^2}{12 \mu Q + ch^3},
$$ 
which stays positive only if $Q > -\frac{ch^3}{12\mu}$.

If $Q < -\frac{ch^3}{12\mu}$, then the form \eqref{au.5} seems to be nonphysical as $u$ is negative in the vicinity of the plates, which means that the fluid should flow against the pressure gradient. This is why we provide a different scenario, in the spirit of the one presented in Section \ref{Sec3}. 

We assume that the fluid near the plates, more specifically in the layers characterized by $y\in (0, h/2 - \overline{h})$ and by $y\in (h/2 + \overline{h}, h)$,  is ``stuck" to the plates and the layer and the wall behave as a rigid body at rest. The precise value of $\overline{h} \in (0, h/2)$ will be specified later. Then for the channel characterized as the set where $y\in (h/2 - \overline{h}, h/2 + \overline{h})$ the solution is as the solution corresponding to no-slip boundary conditions on the plates $y=h/2\pm \overline{h}$. Comparing it with \eqref{au.8} we observe that the solution is of the form 
\begin{equation}
   u(y) = - \frac{c}{2\mu} (h/2 + \overline{h} -y) (y- h/2 + \overline{h})\,. \label{au.9}
\end{equation}
This leads to the following formula for the flow rate:
$$
  Q= - \frac{c}{2\mu} \int_{h/2 - \overline{h}}^{h/2 + \overline{h}} (h/2 + \overline{h} -y) (y- h/2 + \overline{h}) \, \textrm{d}y = - \frac{c}{\mu} \int_{0}^{\overline{h}} \left[ \overline{h}^2 - z^2 \right]\, \textrm{d}z = - \frac{2c}{3\mu} \overline{h}^3,
$$
where we used the substitution $y=h/2 -z$. The above formula then gives 
$$
   \overline{h} = \sqrt[3]{\frac{3\mu Q}{-2 c}}\,.
$$
Note that the requirement $\overline{h} < h/2$ is satisfied due to our assumptions concerning $Q$, $c$, $h$ and $\mu$.

\section{Plane Couette flow} \label{Sec6} 

Similarly as in the previous section, the flow of the Navier-Stokes fluid takes place between two parallel plates located at $y=0$ and $y=h$, the flow being engendered due to the application of a shear stress on the upper plate. Again, $\vec{b}= \vec{0}$ and $\vec{v} = u(y)\vec{i}$, but the pressure $p$ is supposed to be constant. It then follows from \eqref{dod5} that $\diver \T = (\tau', 0, 0)$ where $'$ denotes $\frac{\textrm{d}}{\textrm{d}y}$ through the whole section.

We assume that a shear stress $\tau_{\textrm{app}}$ is applied on the fluid by moving the upper plate in contact with the fluid (this being equal and opposite to the shear stress exerted by the fluid on the plate), i.e.,
\begin{equation}
    \tau(h) = \tau_{\textrm{app}}. \label{c.p1}
\end{equation}
Without loss of generality, we can assume that $\tau_{\textrm{app}}> 0$. Due to $\tau_{\textrm{app}}$, the upper plate moves with the velocity $V$ that can be (easily) measured. 

The governing system of equations \eqref{D0.2} and \eqref{D0.3} reduces to
$$
\tau = \mu u' \quad \textrm{ and } \quad  \tau' = 0\,.
$$ 
Consequently,
\begin{equation}\label{c.p2}
   \tau(y) = C \quad \textrm{ and } \quad u(y) = \frac{C}{\mu}y + D. \qquad\qquad (C,D \textrm{ are constants})
\end{equation}
Using the boundary condition \eqref{c.p1}, we obtain
\begin{equation}\label{c.p3}
 \tau(y) = \tau_{\textrm{app}} \quad \textrm{ and } \quad u(y) = \frac{\tau_{\textrm{app}}}{\mu} y + D\,.
\end{equation}

As in the preceding section, we assume that we know the volumetric flow rate $Q$ and it is a fixed positive constant. Then 
$$
  Q= \int_0^h u(y) \, \textrm{d}y = \frac{\tau_{\textrm{app}}}{2\mu} h^2 + D h
$$
leads to 
$$
  D = \frac{1}{h} \left[Q - \frac{\tau_{\textrm{app}}}{2\mu} h^2\right].
$$
Hence 
\begin{equation} \label{c.p4}
  u (y) = \frac{\tau_{\textrm{app}}}{\mu} y + \frac{1}{h} \left[Q - \frac{\tau_{\textrm{app}}}{2\mu} h^2\right].
\end{equation}
Notice that $u(0) = \frac{1}{h} \left[Q - \frac{\tau_{\textrm{app}}}{2\mu} h^2\right]$. In the rest of this section, we will discuss in detail the following three situations: (i) $Q - \frac{\tau_{\textrm{app}}}{2\mu} h^2= 0$, (ii) $Q - \frac{\tau_{\textrm{app}}}{2\mu} h^2 > 0$, and (iii) $Q - \frac{\tau_{\textrm{app}}}{2\mu} h^2 < 0$.

\textbf{(i)} Let $Q - \frac{\tau_{\textrm{app}}}{2\mu} h^2= 0$. Then it follows from \eqref{c.p4} that $u(0)=0$ and there is ``no-slip" on the lower plate. Looking then  at the upper plate moving with the velocity $V$, we observe that if $u(h)=V$ then there is ``no-slip" also on the upper plate. This (i.e. $u(h)=V$) happens if 
\begin{equation}\label{c.p5}
    V=\frac{\tau_{\textrm{app}} h}{2\mu} + \frac{Q}{h} \qquad \qquad \left(\textrm{Note that } \frac{\tau_{\textrm{app}} h}{2\mu} + \frac{Q}{h} =  \frac{\tau_{\textrm{app}}h}{\mu} = \frac{2Q}{h} \quad\textrm{ as } \quad Q=\frac{\tau_{\textrm{app}} \, h^2}{2\mu}.\right)
\end{equation}

If $V>\frac{\tau_{\textrm{app}} h}{2\mu} + \frac{Q}{h}$, then the fluid slips at the upper plate. If one assume Navier's slip boundary condition \eqref{D0.6}, then in the case of moving boundary 
\begin{equation}
    \label{c.p6} u(h) - V = - \frac{\tau(h)}{\kappa}\,.
\end{equation}
Using \eqref{c.p3} and \eqref{c.p4} we conclude that 
\begin{equation}
    \label{c.p7} \kappa_h:=\kappa = \frac{\tau_{\textrm{app}}}{V - \frac{Q}{h} - \frac{\tau_{\textrm{app}}h}{2\mu}}.
\end{equation}
Note that in this case the fluid exhibits the ``no-slip" on the lower plate and Navier's slip with $\kappa_h$ given in \eqref{c.p7} on the upper plate.

If $V<\frac{\tau_{\textrm{app}} h}{2\mu} + \frac{Q}{h}$, we assume that the fluid ``sticks" to the upper plate and the whole structure moves as a rigid body with the speed $V$ of the upper plate in the layer characterized by $y\in (\overline{h},h)$, where $\overline{h}\in (0,h)$ will be specified later. It means $u(y)=V$ for $y\in (\overline{h},h)$. On $(0, \overline{h})$ the solution is linear (see \eqref{c.p2}) and satisfies the ``no-slip" boundary conditions $u(0) = 0$ and $u(\overline{h})=V$. Hence, $u(y) = V y/\overline{h}$ for $(0,\overline{h})$. The precise value of $\overline{h}$ is determined from the knowledge of the flow rate $Q$. Indeed, 
$$
   Q = \int_0^h u(y) \, \textrm{d} y = \int _0^{\overline{h}} u(y) \, \textrm{d} y + V(h-\overline{h}) = Vh - V\overline{h}/2,
$$
which implies that
$$
  \overline{h} = 2(Vh-Q)/V\,.
$$ 
Note that the requirement that $\overline{h} < h$ is equivalent to 
$V<\frac{\tau_{\textrm{app}} h}{2\mu} + \frac{Q}{h}$, while the requirement that $\overline{h}>0$ follows from $Vh>Q$, which is a natural condition on the data ($Vh$ is the maximal flow rate capacity of the channel if the upper plate moves with the velocity $V$, the lower plate is at the rest and the motion is generated by $\tau_{\textrm{app}}$). All three cases connected with the ``no-slip" boundary condition on the lower plate are drawn in Fig.~\ref{fig2}.

\tikzmath{\l = 110; \h =15; \s=5; \a=20; \txt=23; }
\tikzmath{\hb = \s * \h / \a ;}

\def\couet#1#2#3{%
  \begin{scope} [shift={(0,#1mm)}, x=1mm, y=1mm]
    %
    \draw [] (0,0) coordinate (A) -- ++(\l,0) coordinate (B) -- ++(0,\h) coordinate (C) -- ++(-\l,0) coordinate (D) -- (A) ;
    %
    \draw [>-<] (-2,0) -- node[midway,left] {$h$} ++(0,\h) ;
    \draw [->] (\l-10,\h+2) -- node[midway,above] {$\tau_{app}$} ++(10,0) ;
    %
    \begin{scope}[shift={(10,0)}]
      \draw[very thick](0,0)--(0,\h);
      \ifnum #3 =0
      \draw [densely dotted, thick, domain=0:\h, samples=20] plot({ \a* (\x  / \h  }, \x);
      \foreach \x in {0,1,...,\h}{
        \draw[tips=proper,->, x=1mm, y=1mm] (0, \x) -- ++( { min( \a + \s*#2 , \a* (\x)  / \h)} , 0);
      }
      \fi
      \ifnum #3 >0
      \draw [densely dotted, thick, domain=0:\h, samples=20] plot({ \s + \a* (\x  / \h  }, \x);
      \foreach \x in {0,1,...,\h}{
        \draw[tips=proper,->, x=1mm, y=1mm] (0, \x) -- ++( { min( \a + \s + \s*#2 , \s + \a* (\x)  / \h)} , 0);
      }
      \fi
      \ifnum #3 <0
      \draw [densely dotted, thick, domain=0:\h, samples=20] plot({ -\s + \a* \x  / \h  }, \x);
      \foreach \x in {0,1,...,\h}{
        \draw[tips=proper,->, x=1mm, y=1mm] (0, \x) -- ++( { min( \a -\s + \s*#2 , max(0, -\s + \a* (\x)  / \h))} , 0);
      }
      \fi
      
      \ifnum #3 <0
      \draw [-, gray] (-10,\hb) -- ++(\l,0) ;
      \draw [>-<] (-6,0) -- node[midway,left] {{\tiny $\overline{h}$}} ++(0,\hb) ;
      \fi

      \ifnum #2 = 0
      \draw [->, very thick] (0,\h+2) -- node[midway,above] {$V$} ++(\a+\s*#3,0) ;
      \ifnum #3 < 0 
      \node[fill=black!20, below right] (txt0) at (\txt,\h-2) {\small no-slip};
      \draw[->] (txt0.west) to [out=180, in=-40] (\a-\s,\h-1);
      \node[fill=black!20, above right] (txt0) at (\txt,2) {\small the whole layer behaves as a static rigid body};
      \draw[->] (txt0.west) to [out=180, in=0] (1, \s/2);
      \fi
      \ifnum #3 = 0 
      \node[fill=black!20, below right] (txt0) at (\txt,\h-2) {\small no-slip};
      \draw[->] (txt0.west) to [out=180, in=-40] (\a,\h-1);
      \node[fill=black!20, above right] (txt0) at (\txt,2) {\small no-slip};
      \draw[->] (txt0.west) to [out=180, in=20] (2, 1);
      \fi
      \ifnum #3 > 0 
      \node[fill=black!20, below right] (txt0) at (\txt,\h-2) {\small no-slip};
      \draw[->] (txt0.west) to [out=180, in=-40] (\a+\s,\h-1);
      \node[fill=black!20, above right] (txt0) at (\txt,2) {\small Navier's slip with $\kappa_0$ given by $\eqref{c.p8}$};
      \draw[->] (txt0.west) to [out=180, in=10] (\s+2, 1);
      \fi
      \fi
      
      \ifnum #2 <0
      \draw [->, very thick] (0,\h+2) -- node[midway,above] {$V$} ++(\a-\s+\s*#3,0) ;
      \draw [-, gray] (-10,\h-\hb) -- ++(\l,0) ;
      \ifnum #3 <0
      \draw [>-<] (-4,0) -- node[pos=0.7,left] {{\tiny $\overline{\overline{h}}$}} ++(0,\h-\hb) ;
      \else
      \draw [>-<] (-5,0) -- node[midway,left] {{\tiny $\overline{h}$}} ++(0,\h-\hb) ;
      \fi
      \ifnum #3 < 0 
      \node[fill=black!20, below right] (txt0) at (\txt,\h-2) {\small the whole layer moves with constant velocity $V$};
      \draw[->] (txt0.west) to [out=180, in=-20] (\a-\s-\s,\h-\s/2);
      \node[fill=black!20, above right] (txt0) at (\txt,2) {\small the whole layer behaves as a static rigid body};
      \draw[->] (txt0.west) to [out=180, in=0] (1, \s/2);
      \fi
      \ifnum #3 = 0 
      \node[fill=black!20, below right] (txt0) at (\txt,\h-2) {\small the whole layer moves with constant velocity $V$};
      \draw[->] (txt0.west) to [out=180, in=-20] (\a-\s,\h-1);
      \node[fill=black!20, above right] (txt0) at (\txt,2) {\small no-slip};
      \draw[->] (txt0.west) to [out=180, in=20] (2, 1);
      \fi
      \ifnum #3 > 0 
      \node[fill=black!20, below right] (txt0) at (\txt,\h-2) {\small the whole layer moves with constant velocity $V$};
      \draw[->] (txt0.west) to [out=180, in=-40] (\a,\h-\s/2);
      \node[fill=black!20, above right] (txt0) at (\txt,2) {\small Navier's slip with $\kappa_0$ given by $\eqref{c.p8}$};
      \draw[->] (txt0.west) to [out=180, in=10] (\s+2, 1);
      \fi
      \fi
      
      \ifnum #2 >0
      \draw [->, very thick] (0,\h+2) -- node[midway,above] {$V$} ++(\a+\s+\s*#3,0) ;
      \ifnum #3 < 0 
      \node[fill=black!20, below right] (txt0) at (\txt,\h-2) {\small Navier's slip with $\kappa_h$ given by $\eqref{c.p10}$};
      \draw[->] (txt0.west) to [out=180, in=-30] (\a-\s,\h-1);
      \node[fill=black!20, above right] (txt0) at (\txt,2) {\small the whole layer behaves as a static rigid body};
      \draw[->] (txt0.west) to [out=180, in=0] (1, \s/2);
      \fi
      \ifnum #3 = 0 
      \node[fill=black!20, below right] (txt0) at (\txt,\h-2) {\small  Navier's slip with $\kappa_h$ given by $\eqref{c.p7}$};
      \draw[->] (txt0.west) to [out=180, in=-30] (\a,\h-1);
      \node[fill=black!20, above right] (txt0) at (\txt,2) {\small no-slip};
      \draw[->] (txt0.west) to [out=180, in=20] (2, 1);
      \fi
      \ifnum #3 > 0 
      \node[fill=black!20, below right] (txt0) at (\txt,\h-2) {\small Navier's slip with $\kappa_h$ given by $\eqref{c.p9}$};
      \draw[->] (txt0.west) to [out=180, in=-40] (\a+\s,\h-1);
      \node[fill=black!20, above right] (txt0) at (\txt,2) {\small Navier's slip with $\kappa_0$ given by $\eqref{c.p8}$};
      \draw[->] (txt0.west) to [out=180, in=10] (\s+2, 1);
      \fi
      \fi
      
    \end{scope}
    
    \ifnum #2 =0
    \ifnum #3 =0
    \node[draw, above right] (txtQ) at (0,25) {\begin{minipage}{25mm}\[Q - \frac{ \tau_{app} h^2} {2 \mu} =0 \]\end{minipage}};
    \fi
    
    \ifnum #3 <0
    \node[draw, above right] (txtQ) at (0,25) {\begin{minipage}{25mm}\[Q - \frac{ \tau_{app} h^2} {2 \mu} <0 \]\end{minipage}};
    \fi
    
    \ifnum #3 >0
    \node[draw, above right] (txtQ) at (0,25) {\begin{minipage}{25mm}\[Q - \frac{ \tau_{app} h^2} {2 \mu} >0 \]\end{minipage}};
    \fi
    \fi
    
    \ifnum #3 <0
    \ifnum #2 =0
    \node[draw] (txtV) at (\l+20,\h/2) {\begin{minipage}{25mm}\[V= \frac{\tau_{app}}{\mu} (h-\overline{h}) \]\end{minipage}};
    \fi
    \ifnum #2 >0
    \node[draw] (txtV) at (\l+20,\h/2) {\begin{minipage}{25mm}\[V> \frac{\tau_{app}}{\mu} (h-\overline{h}) \]\end{minipage}};
    \fi
    \ifnum #2 <0
    \node[draw] (txtV) at (\l+20,\h/2) {\begin{minipage}{25mm}\[V< \frac{\tau_{app}}{\mu} (h-\overline{h}) \]\end{minipage}};
    \fi
    \else
    \ifnum #2 =0
    \node[draw] (txtV) at (\l+20,\h/2) {\begin{minipage}{25mm}\[V= \frac{\tau_{app} h}{2 \mu} + \frac{Q}{h}\]\end{minipage}};
    \fi
    \ifnum #2 >0
    \node[draw] (txtV) at (\l+20,\h/2) {\begin{minipage}{25mm}\[V> \frac{\tau_{app} h}{2 \mu} + \frac{Q}{h}\]\end{minipage}};
    \fi
    \ifnum #2 <0
    \node[draw] (txtV) at (\l+20,\h/2) {\begin{minipage}{25mm}\[V< \frac{\tau_{app} h}{2 \mu} + \frac{Q}{h}\]\end{minipage}};
    \fi
    \fi
    
  \end{scope}
}

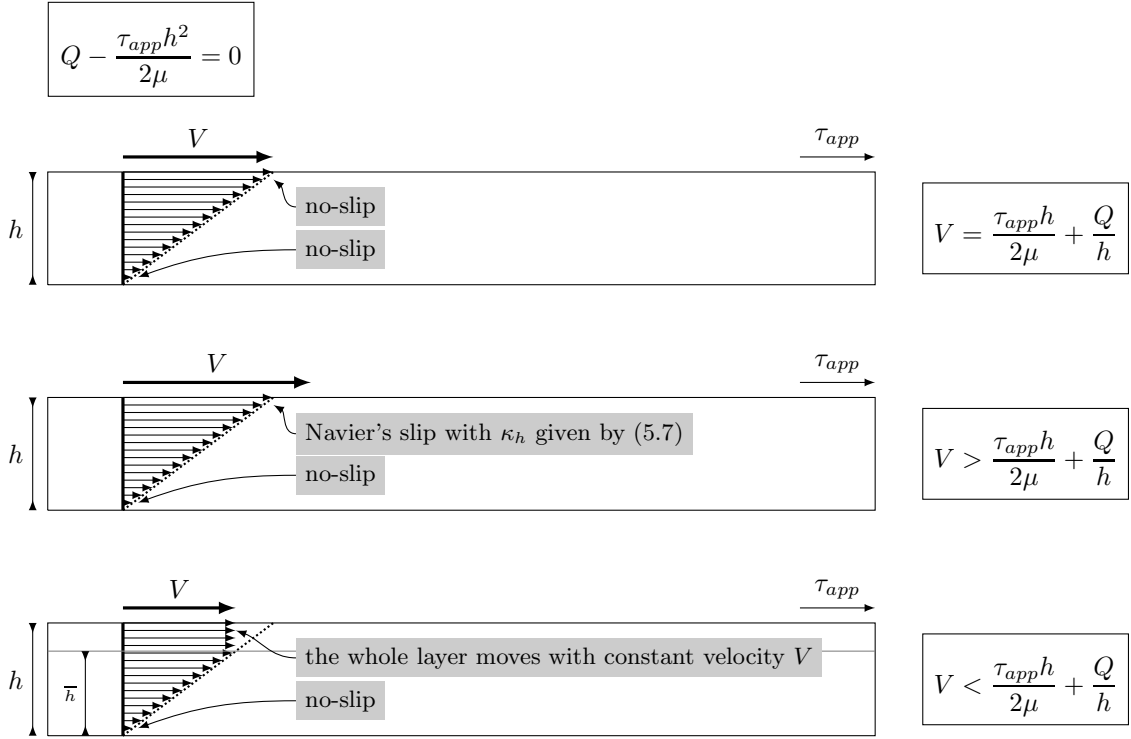
\begin{figure}[ht]
\begin{tikzpicture}
\couet{0}{0}{0}
\couet{-30}{1}{0}
\couet{-60}{-1}{0}
\end{tikzpicture}
\caption{Different flow regimes for plane Couette flows if $Q=\tau_{\textrm{app}}h^2/(2\mu)$. This condition implies the ``no-slip" on the lower plate. Depending on the relation among $V$, $\tau_{\textrm{app}}$, $h$, $\mu$ and $Q$, we identify three different regimes at the upper plate. These regimes are identified by means of (easily) measurable quantities $\mu$, $h$, $\tau_{\textrm{app}}$, $Q$ and $V$.}
\label{fig2}
\end{figure}

\textbf{(ii)} Let $Q - \frac{\tau_{\textrm{app}}}{2\mu} h^2 >  0$. Assuming that the fluid slips according to Navier's slip boundary condition \eqref{D0.6}, which takes the form 
$$
  u(0) = \frac{\tau(0)}{\kappa},
$$
we can fix \emph{positive} $\kappa$. Indeed, 
\begin{equation}
    \label{c.p8}
      \kappa_0:=\kappa = \frac{\tau(0)}{u(0)} = \frac{2\mu \tau_{\textrm{app}}h}{2\mu Q - \tau_{\textrm{app}} h^2}.
\end{equation}

Next, we look at the upper plate. As the constructed solution satisfies 
$$
  u(h) = \frac{\tau_{\textrm{app}}}{\mu} h + \frac{1}{h} \left[Q - \frac{\tau_{\textrm{app}}}{2\mu} h^2\right] = \frac{\tau_{\textrm{app}}}{2\mu} h + \frac{Q}{h},
$$
we observe that if the speed $V$ of the upper plate associated with the applied shear stress $\tau_{\textrm{app}}$ is such that 
$$
  V=\frac{\tau_{\textrm{app}}}{2\mu} h + \frac{Q}{h}\,,
$$
then there is ``no-slip" at the upper plate. Note that in this case we have again different slipping mechanisms at the lower and the upper plate. 

\begin{figure}[ht]
\begin{tikzpicture}
\couet{0}{0}{1}
\couet{-30}{1}{1}
\couet{-60}{-1}{1}
\end{tikzpicture}
\caption{Different flow regimes for plane Couette flows if $Q>\tau_{\textrm{app}}h^2/(2\mu)$. There is Navier's slip on the lower plate with the slip parameter identified by \eqref{c.p8}. There are three possible regimes identified at the upper plate. These regimes are identified by means of experimentally measurable quantities $\mu$, $h$, $\tau_{\textrm{app}}$, $Q$ and $V$.}
\label{fig3}
\end{figure}
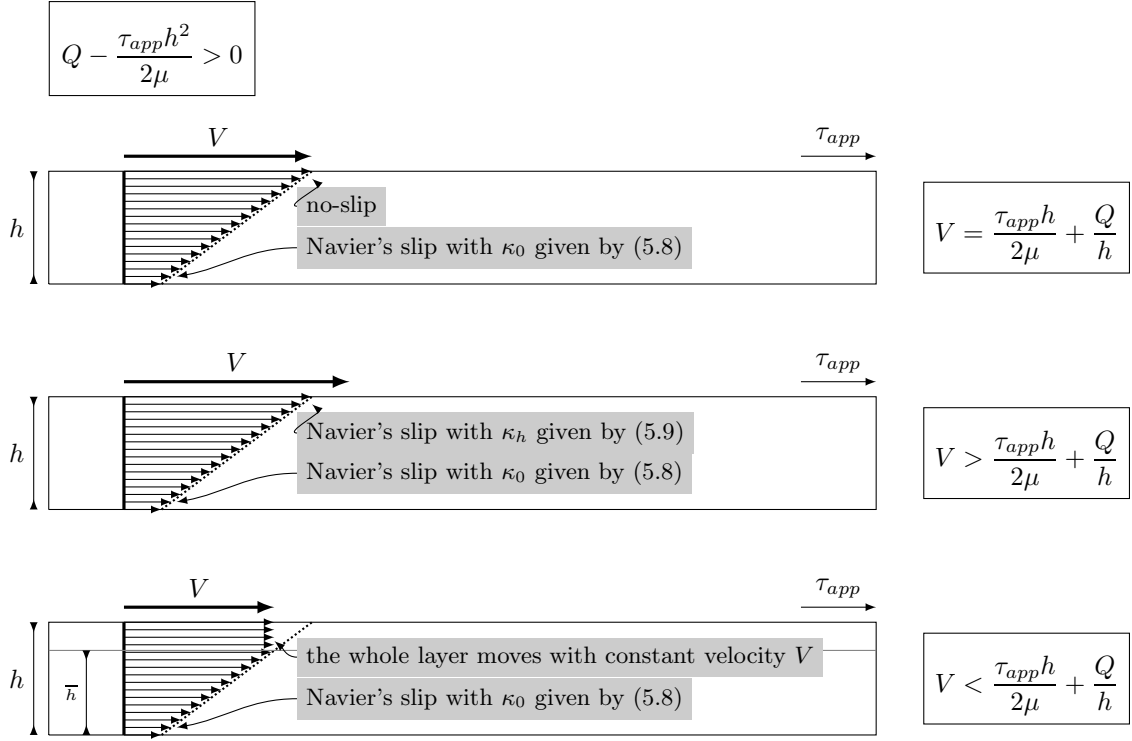

If $V > \frac{\tau_{\textrm{app}}}{2\mu } h + \frac{Q}{h}$, then Navier's slip condition \eqref{c.p6} leads to 
\begin{equation} \label{c.p9}
   \kappa_h = \frac{2\tau_{\textrm{app}} \mu h}{2\mu h V - 2\mu Q - \tau_{\textrm{app}} h^2}, 
\end{equation}
which is due to required condition on $V$, $Q$, $h$, $\mu$ and $\tau_{\textrm{app}}$ positive. Thus, the fluid slips according to Navier'slip boundary condition on the upper an lower plates but the specific value of slip parameter $\kappa$ can be different. In fact, 
$$
\kappa_0 = \kappa_h \quad \textrm{ if } \quad V=2Q/h.
$$
If $V < \frac{\tau_{\textrm{app}}}{2\mu} h + \frac{Q}{h}$, we consider a layer near the upper plate, characterized by $y\in (\overline{h}, h)$, moving as a rigid body with the constant velocity $V$, while on $(0, \overline{h})$ the velocity will be linear satisfying $u(\overline{h})=V$ and $u(0)=\tau(0)/\kappa_0$, where $\kappa_0$ is given in \eqref{c.p8}. It then follows from \eqref{c.p2} that 
$$
u(y) = \frac{V}{\kappa_0\overline{h}+\mu}\left( \kappa_0 y + \mu\right). 
$$
The precise value of $\overline{h}$ can be determined from the requirement that 
$$
Q = \frac{V}{\kappa_0\overline{h}+\mu}\int_0^{\overline{h}} (\kappa_0 y + \mu) \,\textrm{d}y + V(h-\overline{h}),
$$
which leads to a quadratic equation for $\overline{h}$. We draw all three cases associated with Navier's slip boundary condition on the lower plate in Fig.~\ref{fig3}.

\textbf{(iii)} Let $\left[Q - \frac{\tau_{\textrm{app}}}{2\mu} h^2\right] <  0$. In this case we require that a layer in the vicinity of the lower plate, i.e. for $y\in (0,\overline{h})$, is at the rest and the fluid flowing in the region $y\in (\overline{h},h)$ satisfies the condition $u(\overline{h}) = 0$. Here $\overline{h}\in (0, h)$. 
Clearly, $u(y) = 0$ if  $0\le y \le \overline{h}$ and $u(y) = \frac{\tau_{\textrm{app}}}{\mu}(y-\overline{h})$. We determine $\overline{h}$ from the condition 
$$
Q=\int_{\overline{h}}^h u(y) \, \textrm{d} y = \frac{\tau_{\textrm{app}}}{2\mu} (h-\overline{h})^2\,. 
$$
This gives $\overline{h} = h - \sqrt{\frac{2\mu Q}{\tau_{\textrm{app}}}}$. 

Considering the behavior of the solution at the upper plate, we observe that if 
$$
   V = \frac{\tau_{\textrm{app}}}{\mu}(h-\overline{h})
$$
then there in ``no-slip'' on the upper plate. 

If $V > \frac{\tau_{\textrm{app}}}{\mu}(h-\overline{h})$, the assumed Navier's slip boundary condition of the form \eqref{c.p6} leads to 
\begin{equation}\label{c.p10}
    \kappa_h:=\kappa=\frac{\tau_{\textrm{app}}}{V-\frac{\tau_{\textrm{app}}}{\mu}(h - \overline{h})}.
\end{equation}
In this situation, there is a layer near the bottom plate where the fluid is at rest and the fluid slip according to Navier's slip boundary condition, with $\kappa_h$ given in the above formula, on the upper plate. 

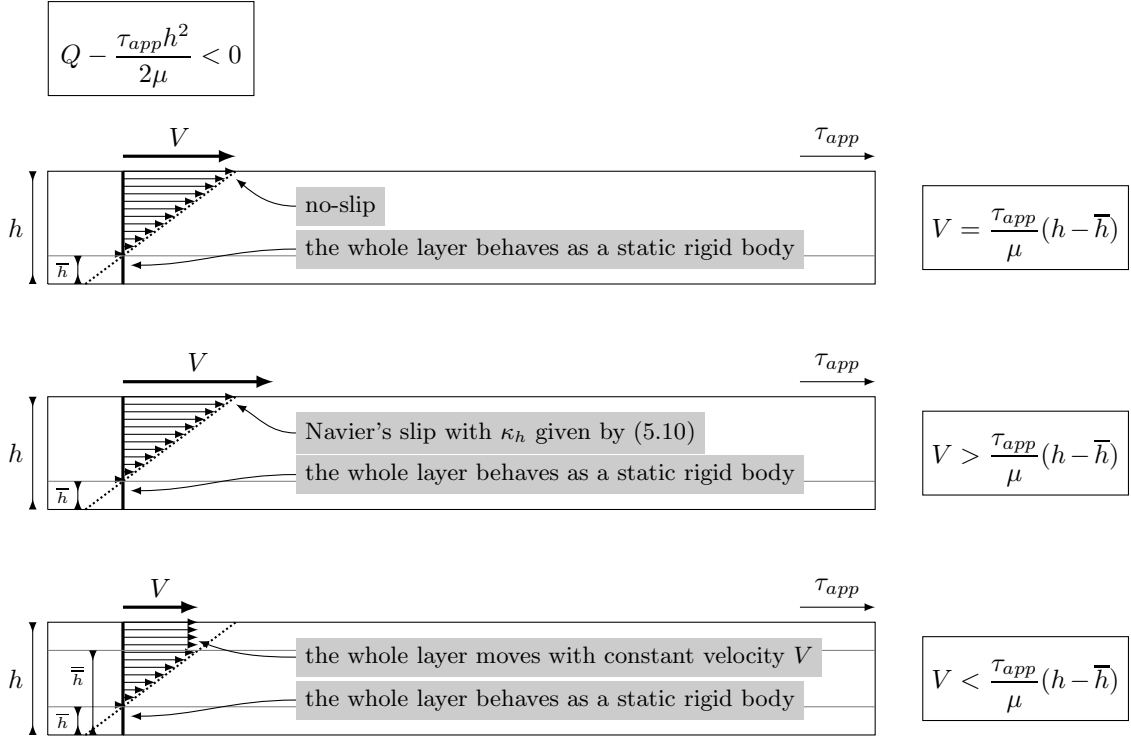
\begin{figure}[ht]
\begin{tikzpicture}
\couet{0}{0}{-1}
\couet{-30}{1}{-1}
\couet{-60}{-1}{-1}
\end{tikzpicture}
\caption{Different flow regimes for plane Couette flows if $Q<\tau_{\textrm{app}}h^2/(2\mu)$. The layer adjacent to the lower plate behaves a rigid body. There are three scenarios identified at the upper plate. These situations are identified by means of (easily) measurable quantities $\mu$, $h$, $\tau_{\textrm{app}}$, $Q$ and $V$.}
\label{fig4}
\end{figure}

If $V < \frac{\tau_{\textrm{app}}}{\mu}(h-\overline{h})$, we assume the existence of the layer of thickness $h-\overline{\overline{h}}$ near the upper plate moving as a rigid body with the constant speed $V$, which is the velocity of the upper plate as measured. The conditions $u(\overline{h})=0$ and $u(\overline{\overline{h}})=V$ lead to 
\begin{equation*}
  u(y)= \begin{cases}
      0&\textrm{ if } y\in (0, \overline{h}), \\
      V (y-\overline{h})/(\overline{\overline{h}} - \overline{h}) &\textrm{ if } y\in (\overline{h}, \overline{\overline{h}}),  \\ V &\textrm{ if } y\in (\overline{\overline{h}}, h).
  \end{cases}
\end{equation*}
The value of $\overline{\overline{h}}$ will be again determined from the condition on the flow rate:
$$
  Q= \int_0^h u(y)\, \textrm{d}y = Vh - V\frac{\overline{h}}{2} - V\frac{\overline{\overline{h}}}{2}
\quad\implies \overline{\overline{h}} = 2h - \overline{h} - \frac{2Q}{V}.
$$
All three cases identified in the case (iii) are drawn in Fig.~\ref{fig4}.


\section{Cylindrical Couette flow} \label{Sec6apul} 

Next, we consider the flow of a Navier-Stokes fluid between concentric cylinders with radii $R_{\textrm{i}}$ (inner cylinder) and $R_{\textrm{o}}$ (outer cylinder), $0<R_{\textrm{i}}<R_{\textrm{o}}$. The Couette flow is characterized by the following conditions:
\begin{equation}
    \vec{v} = v_{\varphi}(r) \vec{e}_{\varphi}= (0,v_{\varphi}(r),0) \quad \text{ and } \quad p= p(r)\,, \qquad \qquad R_{\textrm{i}} < r < R_{\textrm{o}}. \label{jos.0}
\end{equation}
Then, referring to \eqref{A.2}, we have 
\begin{equation}\label{jos.-1}
  \DD = \begin{pmatrix}  0 & \tfrac12 (v_{\varphi}'(r) - v_{\varphi}(r)/r) & 0 \\ 
  \tfrac12 (v_{\varphi}'(r) - v_{\varphi}(r)/r)&0&0 \\ 0 & 0& 0\end{pmatrix}\,.  
\end{equation}
Here and in the rest of this section, $'$ denotes the derivative with respect to $r$. 

Denoting 
$$
   \tau(r):= T_{r\varphi}(r),
$$
it follows from \eqref{jos.-1} and the constitutive equation \eqref{D0.3} (see also \eqref{A.3}) and from the balance equations \eqref{D0.2} with $\vec{b}= \vec{0}$ (see also \eqref{A.1}) that
\begin{align}
v_{\varphi}'(r) - \frac{v_{\varphi}(r)}{r} & = \frac{\tau(r)}{\mu}, \label{jos.1} \\
(r^2 \tau(r))' &= 0, \label{jos.2} \\
\left(\frac{p(r)}{\varrho}\right)'&= \frac{v_{\varphi}^2(r)}{r}. \label{jos.3}
\end{align}

We assume that the outer cylinder is rotating due to the applied torque resulting at the shear stress $\tau_{\textrm{o}}$, i.e.,
\begin{equation}
\tau(R_{\textrm{o}}) = \tau_{\textrm{o}}. \label{jos.4}
\end{equation}

The equation \eqref{jos.2} and the boundary condition \eqref{jos.4} imply that 
\begin{equation}
   \tau(r) = \frac{\tau_{\textrm{o}} R_{\textrm{o}}^2}{r^2}. \label{jos.5}
\end{equation}

Next, using  \eqref{jos.1} we observe from \eqref{jos.5} that
$$
\left(\frac{v_{\varphi}(r)}{r}\right)' = \frac{1}{r} \left(v_{\varphi}'(r) - \frac{v_{\varphi}(r)}{r}\right) = \frac{\tau(r)}{\mu r} = \frac{\tau_{\textrm{o}} R_{\textrm{o}}^2}{\mu r^3},
$$
which leads to
\begin{equation}
    v_{\varphi}(r) = \frac{1}{\mu} \left[D r - \frac{\tau_{\textrm{o}}R_{\textrm{o}}^2}{2r}\right]\,.\label{jos.6}
\end{equation}
To fix the constant $D$, we use the equation for $p$ (see \eqref{jos.3}) and assume that 
\begin{equation}
    p(R_{\textrm{i}})=p_{\textrm{i}} \quad \textrm{ and } \quad p(R_{\textrm{o}})=p_{\textrm{o}}, \label{jos.7}
\end{equation}
where $p_\textrm{i}$ and $p_\textrm{o}$ are given pressures that could be measured by pressure transducers at the outer surface of the inner cylinder and the inner surface of the outer cylinder, respectively. 

Integrating \eqref{jos.3} between $R_i$ and $r$ and using the first condition in \eqref{jos.7} and \eqref{jos.6}, we obtain 
\begin{equation}\label{jos.8}
    \begin{split} 
    \frac{p(r)}{\varrho} &= \frac{p_{\textrm{i}}}{\varrho} +  \int_{R_i}^r \frac{v_{\varphi}^2(s)}{s} \, \textrm{d} s \\ 
    & = \frac{p_{\textrm{i}}}{\varrho} + \frac{1}{\mu^2} \left[ \frac{D^2 r^2}{2} - D \tau_{\textrm{o}} R_0^2 \ln r - \frac{\tau_{\textrm{o}}^2 R_{\textrm{o}}^4}{8r^2} - \frac{D^2 R_{\textrm{i}}^2}{2} + D \tau_{\textrm{o}} R_0^2 \ln R_{\textrm{i}} + \frac{\tau_{\textrm{o}}^2 R_{\textrm{o}}^4}{8 R_{\textrm{i}}^2}\right].
    \end{split}
\end{equation}
Using the second condition in \eqref{jos.7}, we get the quadratic equation for $D$, namely 
$$
   \mu^2\frac{p_{\textrm{o}} - p_{\textrm{i}}}{\varrho} = \frac{D^2}{2}(R_{\textrm{o}}^2 - R_{\textrm{i}}^2) - D \tau_{\textrm{o}}R_{\textrm{o}}^2 \ln\frac{R_{\textrm{o}}}{R_{\textrm{i}}}
   + \frac{\tau_{\textrm{o}}^2 R_{\textrm{o}}^2}{8 R_{\textrm{i}}^2} (R_{\textrm{o}}^2 - R_{\textrm{i}}^2).
$$
Setting $a:=(R_{\textrm{o}}^2 - R_{\textrm{i}}^2)/2$, $b:=-\tau_{\textrm{o}}R_{\textrm{o}}^2 \ln(R_{\textrm{o}}/R_{\textrm{i}})$ and $c:=\tau_{\textrm{o}}^2 R_{\textrm{o}}^2 (R_{\textrm{o}}^2 - R_{\textrm{i}}^2)/(8 R_{\textrm{i}}^2) - \mu^2(p_{\textrm{o}} - p_{\textrm{i}})/\varrho$, this quadratic equation has (one or two) real solutions if and only if $b^2 - 4ac\geq 0$. This gives the restriction on admissible sets of problem parameters $\mu$, $\varrho$, $\tau_{\textrm{o}}$, $R_{\textrm{o}}$, $R_{\textrm{i}}$, $p_{\textrm{o}}$ and $p_{\textrm{i}}$. In what follows, we assume that $D$ is fixed. (We also do not discuss further the possibility that there are two such $D$'s.) 

It follows from \eqref{jos.6} that the angular velocity $\omega(r):=v_{\varphi}(r)/r$ is given through
\begin{equation}
    \omega(r) = \frac{1}{\mu} \left[D  - \frac{\tau_{\textrm{o}}R_{\textrm{o}}^2}{2r^2}\right]\,.\label{jos.9}
\end{equation}
As a consequence of applied torque on the outer cylinder (see \eqref{jos.4}),  the outer cylinder moves with the constant (experimentally measurable) velocity $V_{\textrm{o}}^{\varphi}=\Omega_{\textrm{o}}R_{\textrm{o}}$, where $\Omega_{\textrm{o}}$ stands for the angular speed of the outer cylinder. The inner cylinder is supposed to be at rest, i.e. $V_{\textrm{i}}^{\varphi} = \Omega_{\textrm{i}} R_{\textrm{i}} = 0$. 

With $D=D(p_{\textrm{i}}, p_{\textrm{o}})$ fixed, referring to \eqref{jos.6}, we observe from \eqref{jos.9} that $\omega(R_{\textrm{i}}) = \frac{1}{\mu} \left[D - \frac{\tau_{\textrm{o}}}{2} \frac{R_{\textrm{o}}^2}{R_{\textrm{i}}^2}\right]$. 
We distinguish the following three situations: 
(i) $D = \tfrac{\tau_{\textrm{o}}}{2} \tfrac{R_{\textrm{o}}^2}{R_{\textrm{i}}^2}$, (ii) $D > \tfrac{\tau_{\textrm{o}}}{2} \tfrac{R_{\textrm{o}}^2}{R_{\textrm{i}}^2}$, and (iii) $D < \tfrac{\tau_{\textrm{o}}}{2} \tfrac{R_{\textrm{o}}^2}{R_{\textrm{i}}^2}$.

\textbf{(i)} Let $D = \tfrac{\tau_{\textrm{o}}}{2} \tfrac{R_{\textrm{o}}^2}{R_{\textrm{i}}^2}$. Then $\omega(R_{\textrm{i}}) = 0$ and there is ``no-slip'' on the inner cylinder. Three possibilities are considered at the outer cylinder. 

First, if experimentally measurable parameters $\tau_{\textrm{o}}$, $D$, $\mu$ and $\Omega_{\textrm{o}}$ satisfy
\begin{equation}
    \Omega_{\textrm{o}} = \frac{1}{\mu}\left[ D - \frac{\tau_{\textrm{o}}}{2}\right] \quad \iff \quad 2\mu \Omega_{\textrm{o}} - 2D + \tau_{\textrm{o}} = 0,\label{jos.11}
\end{equation}
then $\omega(R_{\textrm{o}}) = \Omega_{\textrm{o}}$, which means that there is ``no-slip" at the outer cylinder. 

Second, if $2\mu\Omega_{\textrm{o}} - 2D + \tau_{\textrm{o}} > 0$, then, assuming the Navier's slip boundary condition in the form 
$$
   v_{\varphi} (R_{\textrm{o}}) - \Omega_{\textrm{o}} R_{\textrm{o}} = - \frac{1}{\kappa_{\textrm{o}}} \tau(R_{\textrm{o}}) = - \frac{1}{\kappa_{\textrm{o}}} \tau_{\textrm{o}},
$$
we can determine $\kappa_{\textrm{o}}$:  
\begin{equation}
    \label{jos.12}
    \kappa_{\textrm{o}} = \frac{\tau_{\textrm{o}}}{(\Omega_{\textrm{o}} - \omega(R_{\textrm{o}}))R_{\textrm{o}}} = \frac{2\mu\tau_{\textrm{o}}}{(2\mu\Omega_{\textrm{o}} - 2D + \tau_{\textrm{o}})R_{\textrm{o}}}.
\end{equation}

Third, if $2\mu\Omega_{\textrm{o}} - 2D + \tau_{\textrm{o}} < 0$, then we assume that the fluid near the outer cylinder is ``stuck" to the outer cylinder and the whole layer moves with the outer cylinder as the rigid body with the angular velocity $\Omega_{\textrm{0}}$. The fluid responds as a Navier-Stokes fluid in the domain characterized by $r\in (R_{\textrm{i}}, R_{\textrm{o}}-h)$, where $h>0$ will be specified later. In this situation 
\begin{equation}
    \omega(r) = \begin{cases} \frac{1}{\mu} \left[\tilde{D}  - \frac{\tau_{\textrm{o}}R_{\textrm{o}}^2}{2r^2}\right] \quad &\textrm{ if } r\in (R_{\textrm{i}}, R_{\textrm{o}}-h), \\ \Omega_0 &\textrm{ if } r\in (R_{\textrm{o}}-h, R_{\textrm{o}}).\end{cases}\label{jos.13}
\end{equation}
The precise values of $\tilde{D}$ and $h$ are determined from the requirement that 
\begin{equation}\label{jos.14}
\omega(R_{\textrm{o}}-h) = \Omega_{\textrm{o}}.
\end{equation}
and from the knowledge of $p_{\textrm{i}}$, $p_{\textrm{o}}$ and the requirement that $p$ is continuous at $R_{\textrm{o}}-h$. Using \eqref{jos.8}, the latter leads to the condition
\begin{equation}\label{jos.16}
\frac{p_{\textrm{i}}}{\varrho} + \frac{1}{\mu^2} \left[ \frac{ (R_{\textrm{o}}-h)^2 - R_{\textrm{i}}^2} {2} \tilde{D}^2 - \tau_{\textrm{o}} R_0^2 \ln \frac{R_{\textrm{o}}-h}{R_{\textrm{i}}}\, \tilde{D}  - \frac{\tau_{\textrm{o}}^2 R_{\textrm{o}}^4}{8(R_{\textrm{o}}-h)^2}  + \frac{\tau_{\textrm{o}}^2 R_{\textrm{o}}^4}{8 R_{\textrm{i}}^2}\right] = \frac{p_{\textrm{o}}}{\varrho} - \Omega_{\textrm{o}}^2 \frac{R_{\textrm{o}}^2 - (R_{\textrm{o}} - h)^2}{2},
\end{equation}
while \eqref{jos.14} gives
\begin{equation}\label{jos.15}
    R_{\textrm{o}}-h= \sqrt{\frac{\tau_{\textrm{o}}R_{\textrm{o}}^2}{2(\tilde{D} - \mu \Omega_{\textrm{o}})}}.
\end{equation}
By solving the system of two equations \eqref{jos.16} and \eqref{jos.15}, one obtains, under certain additional conditions associated with the quadratic equation \eqref{jos.16}, the values $\tilde{D}$ and $h$ that determine the solution in the form \eqref{jos.13}.

\textbf{(ii)} Let $D > \tfrac{\tau_{\textrm{o}}}{2} \tfrac{R_{\textrm{o}}^2}{R_{\textrm{i}}^2}$. In this case, assuming that the Navier-Stokes fluid slip according to Navier's slip boundary condition, we can determine the value of the slip parameter $\kappa_{\textrm{i}}$. Indeed, from 
$$
   v_{\varphi}(R_{\textrm{i}}) = \frac{1}{\kappa_{\textrm{i}}} \tau(R_{\textrm{i}})
$$
we obtain, using \eqref{jos.5} and \eqref{jos.6}, that 
\begin{equation}
\label{jos.17} 
    \kappa_{\textrm{i}} = \frac{2\mu\tau_{\textrm{o}} R_{\textrm{o}}^2}{(2D R_{\textrm{i}}^2 - \tau_{\textrm{o}} R_{\textrm{o}}^2) R_{\textrm{i}}}.   
\end{equation}
Proceeding in the same way as in the case (i) above, we can distinguish three situations at the outer cylinder. As the procedure is identical, we skip the details and refer the reader to the material discussed above concerning the procedure.

\textbf{(iii)} Let $D < \tfrac{\tau_{\textrm{o}}}{2} \tfrac{R_{\textrm{o}}^2}{R_{\textrm{i}}^2}$. In this case, we need to require that the fluid is ``stuck" to the outer cylinder for $r\in (R_{\textrm{i}}, R_{\textrm{i}}+h)$ while the fluid flows as a Navier-Stokes fluid for $r\in (R_{\textrm{i}}+h, R_{\textrm{o}})$. As $v_{\varphi}(R_{\textrm{i}}+h) = \omega(R_{\textrm{i}}+h) = 0$, the solution is the form 
\begin{equation}
    \omega(r) = \begin{cases} 0 &\textrm{ if } r\in (R_{\textrm{i}}, R_{\textrm{i}}+h) \\ \frac{\tau_{\textrm{o}} R_{\textrm{o}}^2}{2\mu} \left[\frac{1}{(R_{\textrm{i}}+h)^2}  - \frac{1}{r^2}\right] \quad &\textrm{ if } r\in (R_{\textrm{i}}+h, R_{\textrm{o}}).
    \end{cases}\label{jos.18}
\end{equation}
Again, proceeding as in the case (i) above, we can distinguish three situations at the outer cylinder. We skip any further details.

\section{Flow down an inclined plane due to gravity} \label{Sec7}

Let us consider the flow of a Navier-Stokes fluid down an inclined plane. In the coordinate system associated with the inclined plane, the gravitational force in the chosen coordinate system takes the form $\varrho\vec{b} = (\varrho g \sin\theta, -\varrho g \cos\theta, 0)$, where $g$ is the acceleration due to gravity and $\theta$ is the angle of inclination.

We assume that $\vec{v} = u(y) \vec{i}$, $p=p(y)$ and $\tau = \tau(y)$, where again $\tau:=T_{xy}$. The balance of linear momentum \eqref{D0.2} and the constitutive equation \eqref{D0.3} imply that
\begin{align}
    \tau' + \varrho g \sin \theta &= 0\,, \label{1cc} \\
    -p' - \varrho g \cos\theta &= 0\,, \label{1dd} \\
    u' &= \frac{\tau}{\mu},. \label{1ddd}
\end{align}
After integrating \eqref{1cc} we obtain ($\ell$ is a constant)
\begin{equation}
    \tau(y) = -\varrho g (\sin\theta) y + \mu \ell\,.\label{1ee}
\end{equation}
It then follows from \eqref{1ddd} that ($ m$ is a constant)
\begin{equation}
    u(y) = -\frac{\varrho g (\sin\theta)}{2\mu} y^2 + \ell y +  m\,. \label{1ff}
\end{equation}
At the free surface, we assume that $\T\vec{n} = (\tau(h), -p(h), 0)$ vanishes, i.e., 
$$
  \tau(h) = 0\quad \textrm{ and } \quad p(h)=0\,.
$$
The first condition together with \eqref{1ee} yields $\ell= \frac{\varrho g (\sin\theta) h}{\mu}$. Hence,
\eqref{1ff} takes the form 
\begin{equation} \label{1gg} 
 u(y) = \frac{\varrho g \sin\theta}{\mu} \left(h -\frac{y}{2}\right) y +  m\,.  
\end{equation}
Assuming that we know the volumetric flow rate $Q$, we can fix $ m$. As $Q= \int_0^h u(y)\, \textrm{d}y$, we conclude from the above formula for $u$ that 
$$
  Q = \int_0^h \left[ \frac{\varrho g \sin\theta}{\mu} \left(h -\frac{y}{2}\right) y +  m \right]\, \textrm{d}y = \frac{\varrho g (\sin\theta) h^3}{3 \mu} +  m h\,,
$$
which implies that 
\begin{equation} \label{1hh}
m = \frac{Q}{h} - \frac{\varrho g (\sin\theta) h^2}{3 \mu}\,.
\end{equation}
Clearly, if $Q= \frac{\varrho g (\sin\theta) h^3}{3 \mu}$, then $ m = 0$ and consequently $u(0)=0$ and the fluid exhibits ``no-slip'' along the inclined plane.

Let us now assume that $Q \neq \frac{\varrho g (\sin\theta) h^3}{3 \mu}$ and the fluid slips according to \eqref{D0.6} along the inclined plane. In the considered geometrical setting, \eqref{D0.6} simplifies to 
\begin{equation}
    \label{josef4} u(0) = \frac{\tau(0)}{\kappa}.
\end{equation}
Using \eqref{1ee} and \eqref{1gg} together with \eqref{1hh}, this leads to
$$
\kappa = \frac{\tau(0)}{u(0)} = \frac{\mu \ell}{m} = \frac{3\mu \varrho g (\sin\theta) h^2}{3\mu Q - \varrho g (\sin\theta) h^3}.
$$
As $\kappa$ has to be positive, the above formula fixes its value provided that $Q>\varrho g (\sin\theta) h^3/(3\mu)$.

If 
\begin{equation}
    \label{1mm} Q<\varrho g (\sin\theta) h^3/(3\mu),
\end{equation} 
we assume that the fluid is at the rest for all $y\in (0, \overline{h})$ where $\overline{h}\in (0, h)$ will be specified later. Considering the equations \eqref{1cc}--\eqref{1ddd} on $(\overline{h},h)$ together with the boundary conditions
\begin{equation}\label{josef5}
    \tau(h) = 0, \quad p(h)=0\quad  \textrm{ and } \quad u(\overline{h}) = 0, 
\end{equation} 
we obtain 
\begin{equation} \label{1ii} 
 u(y) = \frac{\varrho g \sin\theta}{\mu} \left(h - \frac{\overline{h}}{2} - \frac{y}{2}\right) (y - \overline{h})\,.  
\end{equation}
For given $h$, $\theta$, $\varrho$, $\mu$ and $Q$, we determine $\overline{h}$ from the condition $Q=\int_{\overline{h}}^h u(y)\, \textrm{d} y$ with $u$ given in \eqref{1ii}. This gives\footnote{This is a consequence of the following calculation
\begin{align*}
\frac{\mu Q}{\varrho g \sin\theta} &= \int_{\overline{h}}^h (y-\overline{h})(h-\frac{\overline{h}}{2} - \frac{y}{2})\, \textrm{d}y
 = \int_0^{h-\overline{h}} z (h - \overline{h} - \frac{z}{2})\, \textrm{d}z \\
&=\left[\frac{z^2}{2}\left(h-\overline{h} - \frac{z}{2}\right)\right]_0^{h-\overline{h}} + \frac12\int_0^{h-\overline{h}} \frac{z^2}{2} \, \textrm{d}z = \frac{(h-\overline{h})^3}{4} + \frac{1}{12} [z^3]_0^{h-\overline{h}} = \frac{(h-\overline{h})^3}{3}\,.
\end{align*}
}:
$$
  (h-\overline{h})^3 = \frac{3 \mu Q}{\varrho g \sin\theta} \quad \implies \quad \overline{h} = h - \sqrt[3]{\frac{3 \mu Q}{\varrho g \sin\theta}}. 
$$
Note that $\overline{h}\in (0,h)$ which follows from \eqref{1mm}.

\section{Conclusion}

For unidirectional flows of an incompressible Navier-Stokes fluid with constant viscosity, it is possible to determine the validity of the ``no-slip" on the walls from the knowledge of the viscosity $\mu$, the volumetric flow rate $Q$, the pressure gradient $-c$ and the geometrical dimensions (radius of cylinders or the distance between the boundary plates). This approach, developed in \cite{MKRR2023a} for five different types of flows, thus guarantees the validity of the ``no-slip" if the volumetric flow rate equals to some critical value $Q_{\textrm{crit}}$ (depending on $\mu$, $-c$ and some other relevant macroscopic values). In this study, we investigate the situation when the volumetric flow rate \emph{does not fulfill} the condition guaranteeing the ``no-lip". 

We have found that 
\begin{itemize}
    \item If $Q>Q_{\textrm{crit}}$ and if we assume that the fluid slips past the boundary according to Navier's slip boundary conditions, we have been able to specify the exact value of Navier's slip-parameter $\kappa$. 
    \item If $Q<Q_{\textrm{crit}}$, the fluid cannot meet the Navier's slip boundary condition (as otherwise Navier's slip-parameter $\kappa$ would be negative, which contradicts the second law of thermodynamics and the fluid would flow near the wall in the opposite direction as in the inner part of the pipe). Such a solution does not seem to be in keeping with physics.
    \item If $Q<Q_{\textrm{crit}}$, the following scenario is possible: the fluid near the outer cylinder is ``stuck" to the wall and the layer together with the wall behaves as a rigid body.  In the inner region the Navier-Stokes fluid flows subject to ``no-slip" at the interface. The precise thickness of the layer is determined from the given macroscopic data. By interpreting the layer in the vicinity of the wall as rigid solids, we relax the assumption of the continuity of the stress at the interface between the layer and the fluid flowing in the smaller domain. The scenario described above 
    resembles the situation described in a recent study \cite{Li2020}. Other variants are systematically treated in the case of Poiseuille pipe flow and for plane Couette flow in \cite{Busse2013}.
\end{itemize}



Numerous experimental investigations have been carried out in the past three decades to gain an understanding of the complex mechanisms associated with the slipping of fluids past wall that are rough, hydrophobic, hydrophilic, etc., see the review articles \cite{Neto2005, Priezjev2005, Vinogradova2006, Busse2013, Lee2014} and the several references therein, and the papers \cite{Schnell1956,  Churaevetal1984, H1, H2, Vinogradova1999, Pitet2000, Craigetal2001, Baundryetal2001, Zhu2002, Vinogradova2009,  Penkavova2017}. However, these investigations are not relevant to the kind of flows considered by Du Buat on the basis of which Stokes initially advocated the no-slip boundary condition.


\section{Appendix - governing equations written in the cylindrical coordinates}\label{Sec8}

\noindent
Balance of linear momentum \eqref{D0.2}:
\begin{align}\label{A.1}
\begin{aligned}
    \frac{\partial v_r}{\partial t} + v_r \frac{\partial v_r}{\partial r} + \frac{v_\varphi}{r} \frac{\partial v_r}{\partial \varphi} + v_z \frac{\partial v_r}{\partial z} - \frac{v_{\varphi}^2}{r} &= 
     \frac{1}{\varrho r} \frac{\partial (r T_{rr})}{\partial r} + \frac{1}{\varrho r} \frac{\partial  T_{r\varphi}}{\partial \varphi} + \frac{1}{\varrho} \frac{\partial T_{rz}}{\partial z} - \frac{T_{\varphi\varphi}}{\varrho r} + b_r \\
    \frac{\partial v_\varphi}{\partial t} + v_r \frac{\partial v_\varphi}{\partial r} + \frac{v_\varphi}{r} \frac{\partial v_\varphi}{\partial \varphi} + v_z \frac{\partial v_\varphi}{\partial z} + \frac{v_r v_{\varphi}}{r} &= 
    \frac{1}{\varrho r^2} \frac{\partial (r^2 T_{r\varphi})}{\partial r} + \frac{1}{\varrho r} \frac{\partial T_{\varphi\varphi}}{\partial \varphi} + \frac{1}{\varrho} \frac{\partial T_{\varphi z}}{\partial z} + b_\varphi \\
    \frac{\partial v_z}{\partial t} + v_r \frac{\partial v_z}{\partial r} + \frac{v_\varphi}{r} \frac{\partial v_z}{\partial \varphi} + v_z \frac{\partial v_z}{\partial z} &= 
    \frac{1}{\varrho r} \frac{\partial (r T_{rz})}{\partial r} + \frac{1}{r\varrho} \frac{\partial  T_{\varphi z}}{\partial \varphi} + \frac{1}{\varrho} \frac{\partial T_{zz}}{\partial z} + b_z    
\end{aligned}
\end{align}
The symmetric part of the velocity gradient $\DD = \DD(\vec{v}) = \tfrac12[\nabla \vec{v} + (\nabla \vec{v})^T]$:
\begin{align}\label{A.2}
\DD = \begin{pmatrix}  \frac{\partial v_r}{\partial r}  & \frac12\left[\frac{\partial v_\varphi}{\partial r} + \frac{1}{r} \frac{\partial v_r}{\partial \varphi} - \frac{v_\varphi}{r} \right]& \frac12\left[\frac{\partial v_z}{\partial r} + \frac{\partial v_r}{\partial z} \right]\\
\frac12\left[\frac{\partial v_\varphi}{\partial r} + \frac{1}{r} \frac{\partial v_r}{\partial \varphi} - \frac{v_\varphi}{r} \right]& \frac{1}{r}\frac{\partial v_\varphi}{\partial \varphi} + \frac{v_r}{r} &  \frac12\left[\frac{1}{r} \frac{\partial v_z}{\partial \varphi} + \frac{\partial v_\varphi}{\partial z} \right]\\ \frac12 \left[\frac{\partial v_z}{\partial r} + \frac{\partial v_r}{\partial z} \right] & \frac12 \left[ \frac{1}{r} \frac{\partial v_z}{\partial \varphi} + \frac{\partial v_\varphi}{\partial z} \right] & \frac{\partial v_z}{\partial z} \end{pmatrix}\,.
\end{align}
The constitutive equation \eqref{D0.3}:
\begin{align}\label{A.3}
\frac{1}{\mu} \begin{pmatrix} T_{rr} + p & T_{r\varphi} & T_{rz} \\
T_{r\varphi} & T_{\varphi\varphi} + p &  T_{\varphi z} \\ T_{rz} & T_{\varphi z} & T_{zz}+ p \end{pmatrix} = \begin{pmatrix} 2 \frac{\partial v_r}{\partial r}  & \frac{\partial v_\varphi}{\partial r} + \frac{1}{r} \frac{\partial v_r}{\partial \varphi} - \frac{v_\varphi}{r} & \frac{\partial v_z}{\partial r} + \frac{\partial v_r}{\partial z} \\
\frac{\partial v_\varphi}{\partial r} + \frac{1}{r} \frac{\partial v_r}{\partial \varphi} - \frac{v_\varphi}{r} & \frac{2}{r}\frac{\partial v_\varphi}{\partial \varphi} + \frac{2v_r}{r} &  \frac{1}{r} \frac{\partial v_z}{\partial r} + \frac{\partial v_\varphi}{\partial z} \\ \frac{\partial v_z}{\partial r} + \frac{\partial v_r}{\partial z} & \frac{1}{r} \frac{\partial v_z}{\partial r} + \frac{\partial v_\varphi}{\partial z} & 2 \frac{\partial v_z}{\partial z} \end{pmatrix}\,.
\end{align}
Here, $p:=-\frac13 (T_{rr} + T_{\varphi\varphi} + T_{zz})$. This implies that the left-hand side of the equation is traceless and hence, reading the same
at the right-had side, one has 
$$
   0 = \diver \vec{v} = \operatorname{tr} \DD =  \frac{\partial v_r}{\partial r} + \frac{1}{r}\frac{\partial v_\varphi}{\partial \varphi} + \frac{v_r}{r} + \frac{\partial v_z}{\partial z}\,. 
$$ 

\bigskip\bigskip

\noindent\textbf{Acknowledgment.} J. Málek thanks to Jaroslav Hron, Lenka Košárková, Martin Lanzend\"{o}rfer and Vít Průša for their help and a number of useful discussions.

\providecommand{\bysame}{\leavevmode\hbox to3em{\hrulefill}\thinspace}
\providecommand{\MR}{\relax\ifhmode\unskip\space\fi MR }
\providecommand{\MRhref}[2]{%
  \href{http://www.ams.org/mathscinet-getitem?mr=#1}{#2}
}
\providecommand{\href}[2]{#2}

\end{document}